\documentclass[preprint,12pt]{elsarticle}
\usepackage{color}
\usepackage{graphicx}
\usepackage{amssymb}
\usepackage{multicol}
\usepackage{amsmath}
\usepackage{amsthm}
\usepackage{psfrag}
\usepackage{mathtools}
 \usepackage{eso-pic}
\usepackage{mathtools}

\usepackage{lineno,hyperref}
\usepackage[section]{placeins}
\modulolinenumbers[5]

\begin{document}
\journal{Computers $\&$ Electrical Engineering}
\begin{frontmatter}

\title{Resource Allocation and Relay Selection In Full-Duplex Cooperative Orthogonal Frequency Division Multiple Access Networks}
\author{Jafar Banar and  S. Mohammad Razavizadeh}
\address{School of Electrical Engineering \\ Iran University of Science and Technology (IUST)  \\ J\_ banar@Elec.iust.ac.ir, smrazavi@iust.ac.ir}

\begin{abstract}
This paper is on relay selection in uplink of an in-band full-duplex (IBFD) cooperative cellular network. Assuming an orthogonal frequency division multiple access (OFDMA) cellular network, we develop a relay selection and resource allocation algorithm for this network. The relay selection algorithms select the best relay based on distance between users and signal to interference plus noise ratio (SINR) that operate in amplify and forward (AF) mode. The optimization problem allocates the optimum subcarriers and powers to all users to maximize the average sum-rate of the network. In addition, the constraints of the optimization problem are quality of service (QoS) and transmit power of each user. Simulation results illustrate the good performance of the proposed method.
\end{abstract}

\begin{keyword}
In-Band Full-Duplex (IBFD) \sep Orthogonal Frequency Division Multiple Access (OFDMA) \sep Cooperative Networks \sep Resource Allocation \sep Relay Selection \sep Quality of Service (QoS) \sep Amplify and Forward (AF) \sep Convex Optimization
\end{keyword}

\end{frontmatter}

\section{Introduction}
\label{S:1}
The inevitable high bandwidth requirement in the future cellular networks has led to advent of new technologies. One of this technologies is in-band full-duplex (IBFD) communication. IBFD systems can transmit and receive signals at the same time on the similar frequency band with self-interference (SI) that can be reduced by propagation-domain, analog-circuit-domain, and digital-domain cancellation approaches \cite{1}. If a wireless terminal in a cellular network operates in full-duplex mode, it can potentially double the spectral efficiency of the network relative to half-duplex mode.

In recent years, cooperative communications has been known as one of the main techniques to improve the capacity and coverage that one or more nodes help other nodes to make a better communication \cite{2}-\cite{5}. Cooperative communication is a key enabling technology for optimum spectrum use that applies resource sharing between multiple nodes in the network. This technology has many advantages such as improved coverage, throughput, system capacity, power/battery life and etc. Amplify-and-forward (AF) and decode-and-forward (DF) are the most common relaying strategies in cooperative communications \cite{5}-\cite{8}. In the DF method, the relay re-modulates and retransmits the received noisy signal \cite{5},\cite{6}, while in the AF method relay only amplifies and retransmits its received signal \cite{4}-\cite{6}. The complexity of a DF relay is similar to a base station and is higher than an AF relay \cite{9}. It is also possible to apply orthogonal frequency-division multiplexing (OFDM) in cooperative communication to provide improvement in data rate specifically for broadband wireless networks \cite{4}, \cite{7}, \cite{8}.

A review on previous works in full-duplex networks illustrates that the most of studies have been performed for non-cooperative networks (i.e. when only a direct link between the BS and users exists) \cite{10}. For cooperative communications in full-duplex networks, two modes exist: installing fixed relays and exploiting users as relays \cite{11}-\cite{14}. On the other hand, all of these papers only consider the downlink transmission. \cite{11}, \cite{12} and \cite{15} are based on orthogonal frequency division multiple access (OFDMA) that using subcarrier pairing in the relay network can result in the frequency diversity, and hence, improves system performance \cite{16}. Some papers propose a scenario that there are one source and destination and some relays between them \cite{13}.

One of the other techniques in a cooperative network that can improve the network performance is relay selection. This technique can also be used in full-duplex cellular cooperative networks. There are several relay selection schemes proposed for amplify and forward (AF) and decode and forward (DF) networks \cite{17}-\cite{18}. The proposed relay selection methods in \cite{17} are based on channel coefficients, and methods in \cite{18} are based on distances between nodes. 

In this paper, we investigate the effect of relay selection schemes on the total sum-rate of the full-duplex cellular cooperative OFDMA networks. The cooperative network is based on AF relay nodes that are selected from the users that are close to the BS. Afterward, we optimize the allocation of powers to all users and then select the best relay for each of far users. After relay selection, the OFDMA subcarriers are optimally allocated to all users. 

Finally, we have a total sum-rate maximization problem based on relay selection in which the power and quality of service for each user are our constraints. We prove that this problem can be converted to a convex optimization problem and then, we can solve it with numerical methods. Our simulation results demonstrate the performance of our system model and its effect on the total sum rate of the network.

The contributions of this paper are as follows. We investigate uplink of an IBFD-OFDMA cooperative cellular network. In this network, users that are not in the BS's coverage area can communicate with the BS by help of the relays. In our system model, all nodes are IBFD and we analyze the SI effect in the BS on the sum-rate. We consider the large-scale path loss based on users' location. In the subcarrier assignment, we applied the Munkres algorithm \cite{19} to trust that in each time slot, each subcarrier assigned to only one user. We propose relay selection algorithm based on channel coefficient and location of users to choose the best relay to connect users and the BS. Our optimization problem is a sum-rate maximization problem of subcarrier and power allocation with power and quality of service constraints for each user. This problem is a mixed integer nonlinear program (MINLP) that is non-convex. We relax and convert it to a convex problem and then, solve it with numerical methods.

The remainder of this paper is organized as follows. The system model is introduced in Section II. Relay selection schemes are introduced in Section III. Problem formulation for sum-rate maximization is introduced in Section IV. The simulation results are demonstrated in Section V. Finally, Section VI concludes the paper.

\section{System Model}\label{S:2}
\begin{figure}[h]
\centering
\includegraphics[width=5in]{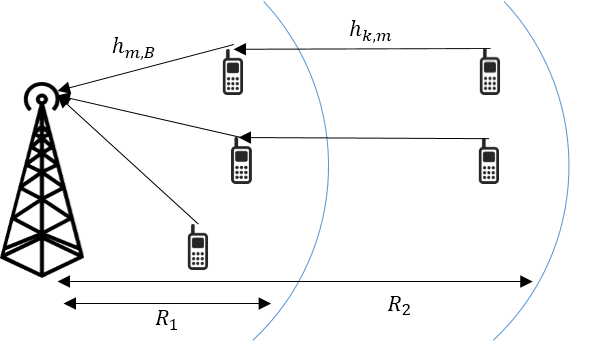}
\caption{System model}
\label{f1}
\end{figure}

We consider uplink of a single cell network which consists of a BS at the center of the cell and two groups of users around it, as depicted in Fig $\ref{f1}$. All users and the BS operate in full-duplex mode. Our OFDMA cellular network's total bandwidth is $NW$, where $N$ is the number of subcarriers and $W$ is the bandwidth of each subcarrier that is the same for all subcarriers. The users of the first group that have far distance to the BS, to communicate with the BS transmit their data to users of the second group that have short distance to the BS. The users of the second group for relaying the received signal from far users that are not in the BS coverage area to the BS, use amplify and forward relaying strategy. The number of users is $K1+K2$ that $K1$ is the number of users in the first group and $K2$ is the number of users in the second group.

In our OFDMA relay selecting system, we choose the best relay to transmit the received signal from the users of the first group to the BS. In general, we need two time slots for data transmission from the users of the first group to the BS in cooperative mode. Any user of the second group that is not selected as relay can send its data to the BS in both time slots in non-cooperative mode. Therefore, the users can be in either cooperative or non-cooperative modes. We also assumed that all users have a single antenna. We assume that in our system model there are only two hops, from the users to the relays and from the relays to the BS. Consequently, in first time slot, all users in the first group transmit their data to the relays (second group users) and also, all users in the second group transmit their data to the BS. In the second time slot, users that work as relay, transmit the amplified data to the BS (cooperative mode) and users that do not work as relay, transmit their data to the BS (non-cooperative mode). Then, relays have to work in IBFD mode to transmit and receive simultaneously in the first time slot. In addition, our relays are not fixed and we use proactive scheme for relay selection that it selects a single relay before data transmission \cite{20}. Our relays operate in non-regenerative mode.

We assume that the BS is at the origin, users that can be used as relays are in the inner boundary with radius $R1$ and far users are between the inner and the outer boundary with radius $R2$. The location of the users can be modeled as the distance $r$ away from the BS and in the random direction of angle $\theta$. Then, closer users to the BS (relays) are at the location $x_r = (r_1,\theta_1)$ and farther of them (users) are at the location $x_u = (r_2,\theta_2)$. We consider the small-scale Rayleigh fading and the large-scale path loss to have more realistic propagation model. We use the standard singular path loss model for the large-scale path loss, $l(x_1,x_2) = ||x_1-x_2||^{-\alpha}$, where $2\leq\alpha\leq6$ \cite{21}. In this equation, $\alpha$ and $||x_1-x_2||$ illustrate the path loss exponent and the Euclidean distance between two nodes, respectively. Therefore, when $x_1$ is a relay and $x_2$ is the BS, the large-scale path loss can be modeled as $l(x_r,0) = l(x_r) = (r_1^2)^{-\frac{\alpha}{2}}$, where $0$ illustrates the origin. Also, the large-scale path loss when, $x_1$ is a user and $x_2$ is a relay can be modeled as $l(x_r,x_u) = (r_1^2+r_2^2-2r_1r_2cos(\theta_1-\theta_2))^{-\frac{\alpha}{2}}$. All distances between the BS and users that are located in the inner boundary and between the inner and the outer boundary assumed to have uniform distribution with probability distribution function (pdf) of

\begin{equation}
\begin{gathered}
\label{e1}
f_{r_1}(r) = \frac{1}{R_1},  	 0\leq r < R_1, \\
f_{r_2}(r) = \frac{1}{R_2-R_1},  	 R_1\leq r < R_2.
\end{gathered}
\end{equation}

In addition, the angle of users assumed to have uniform distribution
\begin{equation}
\label{e2}
f_{\theta}(\theta) = \frac{1}{2\pi},  	 0\leq \theta < 2\pi. 
\end{equation}

\subsection{Signal Model}
The received signal at the relay in the $i$th subcarrier and the first time slot in cooperative mode is

\begin{equation}
\label{e3}
Y_{k,m}^{i,(C,1)} =  \sqrt{P_{k,m}^{i,(C,1)} l(x_r,x_u) |h_{k,m}^i|^2} X_{k,m}^i + Z_{k,m}^i , \\
\end{equation}
\\
Where, $h_{k,m}^i$ denotes the channel coefficients matrix (small scale fading) between the $k$th user in the first group that are far from the BS to the $m$th user in the second group that are close to the BS in the $i$th subcarrier. In general, in this paper, $i$ and $j$ are the subcarrier superscripts in the first and second time slots, respectively. $P_{k,m}^{i,(C,1)}$ is the transmission power of the $k$th user in the first group in the $i$th subcarrier and in the first time slot to the $m$th user in the second group in cooperative mode. In addition, $Z_{k,m}^i$ is the additive white Gaussian noise (AWGN) with variance of $N_0$ received by the $m$th user in the second group in the $i$th subcarrier. Also, $X_{k,m}^i$ is the transmitted symbol from the $k$th user in the first group for the $m$th user in the second group in the $i$th subcarrier. The received signal at the BS in the $j$th subcarrier and in the second time slot in the cooperative mode is

\begin{equation}
\begin{split}
\label{e4}
Y_{m,B}^{j,(C,2)} &= \sqrt{P_{m,B}^{j,(C,2)} l(x_r) |g_{m,B}^j|^2} G Y_{k,m}^{i,(C,1)} \\
&\qquad + \sqrt{P_{B,m}^{j,(C,2)} |H_{SI}^j|^2} X_{B,m}^j + Z_{m,B}^j\\
&= \sqrt{P_{m,B}^{j,(C,2)} P_{k,m}^{i,(C,1)} l(x_r) |g_{m,B}^j|^2 l(x_r,x_u) |h_{k,m}^i|^2} G X_{k,m}^i \\
&\qquad+ \sqrt{P_{B,m}^{j,(C,2)} |H_{SI}^j|^2} X_{B,m}^j \\ 
&\qquad + \sqrt{P_{m,B}^{j,(C,2)} l(x_r) |g_{m,B}^j|^2} G Z_{k,m}^i+ Z_{m,B}^j ,\\ 
\end{split}
\end{equation}

Where, $g_{m,B}^j$ denotes the channel coefficients matrix between the $m$th user in the second group to the BS in the $j$th subcarrier. The self-interference of the BS transmit and receive antennas in the $j$th subcarrier is denoted by the matrix $H_{SI}^j$. $P_{B,m}^{j,(C,2)}$ represents the transmit power from the BS in cooperative mode to the $m$th user in the second group in the $j$th subcarrier and in the second time slot in the downlink. We consider that the noise power is equal to $N_0W$ in each subcarrier for all receivers. In addition, $G$ represents the received signal amplification factor at the AF relays \cite{5} and it is equal to 

\begin{equation}
\begin{aligned}
\label{e5}
G =\frac{1}{ \sqrt{P_{k,m}^{i,(C,1)} l(x_r,x_u) |h_{k,m}^i|^2+N_0W}}.\\
\end{aligned}
\end{equation}

In the non-cooperative mode, the received signal at the BS in the first and second time slots are, respectively, represented as

\begin{equation}
\begin{split}
\label{e6}
Y_{m,B}^{i,(NC,1)} &= \sqrt{P_{m,B}^{i,(NC,1)} l(x_r) |g_{m,B}^i|^2} X_{m,B}^i \\
&\qquad+ \sqrt{P_{B,m}^{i,(NC,1)} |H_{SI}^i|^2} X_{B,m}^i + Z_{m,B}^i ,\\
\end{split}
\end{equation}

\begin{equation}
\begin{split}
\label{e7}
Y_{m,B}^{j,(NC,2)} &= \sqrt{P_{m,B}^{j,(NC,2)} l(x_r) |g_{m,B}^j|^2} X_{m,B}^j \\
&\qquad+ \sqrt{P_{B,m}^{j,(NC,2)} |H_{SI}^j|^2} X_{B,m}^j  + Z_{m,B}^j .\\
\end{split}
\end{equation}

The SINR for cooperative mode is derived as

\begin{equation}
\begin{aligned}
\label{e8}
\Gamma_{UC} = \frac{P_{m,B}^{j,(C,2)} P_{k,m}^{i,(C,1)} \alpha_{k,m}^i \beta_{m,B}^j} {1 + P_{m,B}^{j,(C,2)} \beta_{m,B}^j + P_{k,m}^{i,(C,1)}\alpha_{k,m}^i + P_{B,m}^{j,(C,2)}\gamma_{SI}^j + P_{B,m}^{j,(C,2)}P_{k,m}^{i,(C,1)}\gamma_{SI}^j\alpha_{k,m}^i} .\\
\end{aligned}
\end{equation}

In the denominator of $\eqref{e8}$, we can ignore the terms, $1$ and $P_{B,m}^{j,(C,2)}\gamma_{SI}^j$.
Also, the SINR for the non-cooperative mode is as follows
\begin{equation}
\begin{aligned}
\label{e9}
\Gamma_{UNC1} &= \frac{P_{m,B}^{i,(NC,1)} \beta_{m,B}^i} {1 + P_{B,m}^{i,(NC,1)} \gamma_{SI}^i}, &\\
\Gamma_{UNC2} &= \frac{P_{m,B}^{j,(NC,2)} \beta_{m,B}^j} {1 + P_{B,m}^{j,(NC,2)} \gamma_{SI}^j}, \\
\end{aligned}
\end{equation}
\\
where,
\begin{equation}
\begin{gathered}
\label{e10}
\alpha^i_{k,m} = \frac{l(x_r,x_u) |h^i_{k,m}|^2}{N_0W} \;, \\
\beta^i_{m,B} = \frac{l(x_r) |g^i_{m,B}|^2}{N_0W} \; , \; T = 1, \\
\beta^j_{m,B} = \frac{l(x_r) |g^j_{m,B}|^2}{N_0W}\; , \; T = 2, \\
\gamma^j_{SI} = \frac{|H^j_{SI}|^2}{N_0W}\;,
\end{gathered}
\end{equation}
\\
where, $T$ represents the time slot.
Using the above SINRs, sum-rate of the network in the cooperative and non-cooperative modes are derived as follows\\
\begin{subequations}
\begin{gather}
\label{e11}
R_{k,m}^{i,j,(C)} = \frac{1}{2} \log_2(1+\Gamma_{UC}),\\
R_{m}^{i,(NC,l)} = \frac{1}{2} \log_2(1+\Gamma_{UNCl}), l=1,2.
\end{gather}
\end{subequations}
\\
Total sum-rate is equal to

\begin{equation}
\begin{split}
\label{e12}
R_T &= \sum_{m=1}^{K2} \sum_{k=1}^{K1} \sum_{i=1}^N \sum_{j=1}^N \rho_{k,m}^{i,j} R_{k,m}^{i,j,(C)}  + \sum_{m=1}^{K2} \sum_{i=1}^N \sigma_{m}^{i,(1)} R_{m}^{i,(NC,1)} \\
&\qquad + \sum_{m=1}^{K2} \sum_{j=1}^N \sigma_{m}^{j,(2)} R_{m}^{j,(NC,2)} .\\
\end{split}
\end{equation}
\\
In $\eqref{e12}$, $\rho$ and $\sigma$ are subcarrier indicators, that are equal to 0 or 1. If the $k$th user is transmitting to the BS with the assistance of the $m$th relay in subcarrier pair (i,j), $\rho_{k,m}^{i,j}$ is one, otherwise it is zero. Also, if the $m$th user in the second group is transmitting to the BS in the $i$th subcarrier in the first time slot, $\sigma_m^{i,(1)}$ is one, otherwise it is zero. In addition, in the second time slot, if the $m$th user in the second group is transmitting to the BS in the $j$th subcarrier, $\sigma_m^{j,(2)}$ is one, otherwise it is zero.

\section{Relay Selection}\label{S:3}
In this section, we compare six schemes for relay selection in IBFD OFDMA relaying system. In this paper, users that are close to the BS can act as relays and must be appropriately selected to help far users that are not in the BS coverage area. In contrast to other techniques like SCP method \cite{22} that separately selects the subcarriers with highest signal-to-noise in each hop, in our proposed scheme, we use the criteria that jointly consider both hops. We assume that the distance between farther users and closer users (i.e. relays) and the distance between the closer users and the BS are denoted by $d_u$ and $d_r$, respectively. Then, we use the following schemes for relay selection. 

\subsection{Best SINR Relay Selection}
In this scheme, the best relay in each subcarrier is selected based on the maximum SINR at the BS. This scheme is proposed in \cite{17}. With consideration of the SI effect, we have
\begin{subequations}
\begin{gather}
\label{e13}
r_s = arg \; max(SINR) \\
SINR = \frac{P_{k,m}^{i,(C,1)} P_{m,B}^{j,(C,2)} \alpha^i_{k,m} \beta^j_{m,B}}{P_{k,m}^{i,(C,1)} \alpha^i_{k,m} + P_{m,B}^{j,(C,2)} \beta^j_{m,B} + P_{k,m}^{i,(C,1)} P_{B,m}^{j,(C,2)} \gamma^j_{SI} \alpha^i_{k,m}}
\end{gather}
\end{subequations}
\\
and without consideration of SI effect, we have
\begin{subequations}
\begin{gather}
\label{e14}
r_s = arg \; max(SINR) \\
SINR = \frac{P_{k,m}^{i,(C,1)} P_{m,B}^{j,(C,2)} \alpha^i_{k,m} \beta^j_{m,B}}{P_{k,m}^{i,(C,1)} \alpha^i_{k,m} + P_{m,B}^{j,(C,2)} \beta^j_{m,B}}
\end{gather}
\end{subequations}

\subsection{Best Harmonic Mean Selection}
The best harmonic mean selection is proposed in \cite{17}, that relay with largest harmonic mean are selected as the best relay for data transmission. The selection function is as follows 
\begin{subequations}
\begin{gather}
\label{e15}
r_s = arg \; max(A) \\
A = \frac{2}{(P_{k,m}^{i,(C,1)} \alpha^i_{k,m})^{-1}  + (P_{m,B}^{j,(C,2)} \beta^j_{m,B})^{-1}}
\end{gather}
\end{subequations}

\subsection{Shortest Distance to Relay Selection}
In this scheme, the relay that has the shortest distance to the far users can be selected as the best relay for data transmission \cite{18}. It is assumed that closer users to the BS (relays) are at the location $x_r = (r_1,\theta_1)$ and farther of them (users) are at the location $x_u = (r_2,\theta_2)$. Let, $d_r$ and $d_u$ be the distance between the BS and the relays and the distance between the relays and the users, respectively.The selection function is as follows 
\begin{equation}
\label{e16}
r_s = arg \; min(d_{u})
\end{equation}

\subsection{Shortest Total Distance Selection}
In this scheme, the relay that has the shortest total distance between the user and the BS can be selected as the best relay for data transmission \cite{18}. The selection function is as follows 
\begin{equation}
\label{e17}
r_s = arg \; min(d_{u}+d_{r})
\end{equation}

\subsection{Least Longest Hop Selection}
In this scheme, the relay that has the longest distance of the user and the relay or the relay and the BS can be selected as the best relay for data transmission \cite{18}. The selection function is as follows 
\begin{equation}
\label{e18}
r_s = arg \; min(max(d_{u},d_{r}))
\end{equation}

\subsection{Shortest Distance of Second Hop Selection}
In this scheme, the relay that has the shortest distance to the BS can be selected as the best relay for data transmission \cite{18}. The selection function is as follows 
\begin{equation}
\label{e19}
r_s = arg \; min(d_{r})
\end{equation}
\\
After relay selection, in each subcarrier pair (i,j), the communication mode (cooperative or non-cooperative) based on data rate can be specified. Then, we have an $N\times N$ matrix that its row and column illustrate the subcarriers in the first and second time slots, respectively. In each subcarrier pair (i,j), if the communication mode is cooperative, we use the cooperative data rate in the matrix, and if the communication mode is non-cooperative mode, we use the sum of two data rates in the first and second time slots. Afterward, we apply the Munkres algorithm to this matrix for subcarrier assignment.

\section{Optimization Problem Formulation}\label{S:4}
Our optimization problem based on power and QoS constraints can be written as

\begin{subequations}
\begin{align}
%\label{e20}
   &\max_{P,\rho,\sigma}
   \begin{aligned}[t]
	R_T &= \sum_{m=1}^{K2} \sum_{k=1}^{K1} \sum_{i=1}^N \sum_{j=1}^N \rho_{k,m}^{i,j} R_{k,m}^{i,j,(C)}  + \sum_{m=1}^{K2} 						\sum_{i=1}^N \sigma_{m}^{i,(1)} R_{m}^{i,(NC,1)} \\
				&\qquad + \sum_{m=1}^{K2} \sum_{j=1}^N \sigma_{m}^{j,(2)} R_{m}^{j,(NC,2)}   
   \end{aligned} \notag \\
   &\text{subject to} \notag \\
   &\qquad \rho_{k,m}^{i,j},\sigma_{m}^{i,(1)},\sigma_{m}^{j,(2)} \in{(0,1)}, \quad \forall{k,m,i,j} \label{subeqn:a} \\
   &\qquad \sum_{m=1}^{K2} \sum_{k=1}^{K1} \sum_{j=1}^N \rho_{k,m}^{i,j} + \sum_{m=1}^{K2} \sigma_{m}^{i,(1)} = 1, \quad  \forall{i} \label{subeqn:b} \\
   &\qquad \sum_{m=1}^{K2} \sum_{k=1}^{K1} \sum_{i=1}^N \rho_{k,m}^{i,j} + \sum_{m=1}^{K2} \sigma_{m}^{j,(2)} = 1, \quad  \forall{j} \label{subeqn:c} \\
   &\qquad P_{k,m}^{i,(C,1)}, P_{m,B}^{j,(C,2)}, P_{m,B}^{i,(NC,1)}, P_{m,B}^{j,(NC,2)} \geq 0, \quad  \forall{k,m,i,j} \label{subeqn:d} \\
   &\qquad P_{B,m}^{j,(C,2)}, P_{B,m}^{i,(NC,1)}, P_{B,m}^{j,(NC,2)} \geq 0, \quad  \forall{k,m,i,j} \label{subeqn:e} \\
   &\qquad \sum_{k=1}^{K1} \sum_{j=1}^N \rho_{k,m}^{i,j} (P_{k,m}^{i,(C,1)}+P_{m,B}^{j,(C,2)}) \leq Pmax^C_{k,m} \label{subeqn:f} \\
   &\qquad \sum_{l=1}^N \sigma_{m}^{l,(1)} P_{m,B}^{l,(NC,1)} \leq Pmax^{NC}_{m}, \quad  l = i, j  \label{subeqn:g} \\
   &\qquad \sum_{m=1}^{K2} \sum_{j=1}^N \sum_{i=1}^N \rho_{k,m}^{i,j} R_{k,m}^{i,j,(C)} \geq R_{k}^{(C),min} \label{subeqn:h} \\
   &\qquad \sum_{l=1}^N \sigma_{m}^{l,(1)} R_{m}^{l,(NC,1)} \geq R_{m}^{(NC),min}, \quad  l = i, j  \label{subeqn:i}
\end{align}
\end{subequations}
\\
In the above obtimization problem, $\eqref{subeqn:a}$ to $\eqref{subeqn:c}$ are related to subcarrier allocation. $\sigma_{m}^{i,(1)}$ and $\sigma_{m}^{j,(2)}$ in the $\eqref{subeqn:a}$ show the subcarrier assigning to the users in the non-cooperative mode and $\rho_{k,m}^{i,j}$ shows the subcarrier assigning to the users in the cooperative mode.
$\eqref{subeqn:b}$ and $\eqref{subeqn:c}$ represent the user limitation to experience one type of communication in each time slot. At each time slot only one user can communicate with relays or the BS. $\eqref{subeqn:d}$ to $\eqref{subeqn:g}$ show the positivity of powers and their boundary in cooperative and non-cooperative modes. The constraints $\eqref{subeqn:h}$ and $\eqref{subeqn:i}$ show that the data rate of each user must be more than a minimum requirement in each mode. The constraints $\eqref{subeqn:f}$ to $\eqref{subeqn:i}$ are the quality of service (QoS) constraints.

\subsection{Solution Approach}
In this section, we prove the convexity of our optimization problem in equation (20). First, we investigate the concavity of the uplink data rate of the cooperative mode. Later we will discuss about the non-cooperative mode. To prove the concavity of data rate $R_c = \log(f)$, we need to show that the Hessian matrix of $f = 1+\Gamma_{UC}$ is negative semidefinite. That is, the eigenvalues of this Hessian matrix must be non-positive. Since the logarithm function is a concave and increasing function, if $f$ is concave, $R$ is concave too. Since the first term in $\eqref{e12}$ is concave (by properties of perspective operation), by the notations introduced in the table $\eqref{t1}$, we formulate the denominator of $\eqref{e8}$ as $a*c*x+b*y$. 

\begin{table}[h]
\caption{The notations that used in the Hessian matrix calculation}
\centering
\label{t1}
\begin{tabular}{|c|c|} 
\hline
 $P_{k,m} = x$ &$P_{m,B} = y$ \\
\hline
$\alpha_{k,m} = a$ &$\beta_{m,B} = b$ \\
\hline
$P_{B,m} \gamma_{SI} = c$ &$P_{B,m} = z$ \\
\hline
\end{tabular}
\end{table}

Then, the Hessian matrix of $f$ is as follows
\begin{subequations}
\begin{align}
 \begin{pmatrix}
  \frac{2a^3bc^2xy}{(by+acx)^3}-\frac{2a^2bcy}{\sigma_2} & \sigma_1, \label{subeqn:e21} \\
  \sigma_1 & \frac{2ab^3xy}{(by+acx)^3}-\frac{2ab^2x}{\sigma_2} 
 \end{pmatrix},  
\end{align}\\
\qquad \text{where,}  \\ 
\begin{gather} 
  \sigma_1 = \frac{ab}{by+acx}-\frac{ab^2y}{\sigma_2}-\frac{a^2bcx}{\sigma_2}+\frac{2a^2b^2cxy}{(by+acx)^3},  \\
 \sigma_2 = (by+acx)^2 . 
\end{gather}
\end{subequations}

The determinant of the matrix $\eqref{subeqn:e21}$ is zero, and its eigenvalues are as follows\\
\begin{equation}
\left\{ 0 , -\frac{2ca^2b^2x^2+2ca^2b^2y^2}{a^3c^3x^3+3a^2bc^2x^2y+3ab^2cxy^2+b^3y^3} \right\}. \\
\label{e22}
\end{equation}

The eigenvalues calculated above, are non-positive. It is proved that our function is concave and negative semidefinite. Now, we study the convexity of the uplink data rate equation of non-cooperative mode. We need to calculate the Hessian matrix of the data rate function in this mode to prove that it is a concave function. That is,\\
\begin{equation}
\Bigg( -\frac{b^2}{ln(2)(\frac{bx}{cz+1}+1)^2(cz+1)^2} \Bigg) .\\
\label{e23}
\end{equation}

The determinant of the above Hessian matrix is zero. Now, we calculate its eigenvalues\\
\begin{equation}
\left\{ 0,-\frac{b^2}{ln(2)(bx+cz+1)^2} \right\} .\\
\label{e24}
\end{equation}

It can be seen that the eigenvalues are non-positive. Therefore, it is proved that our function is concave. Now, we can use standard numerical solution methods for convex optimization problems to solve this problem.

\section{Numerical Results}\label{S:5}
In this section, we represent our simulation results that have been performed to evaluate the performance of the proposed schemes. We consider a single cell and a BS that is located at its center. We also have two groups of users that are uniformly distributed around the BS. Users that are placed in a longer distance from the BS are the first group and users in a shorter distance are the second group. All channel coefficients between users, between user and the BS, and also self-interference channels are independent and identically distributed complex Gaussian random variables with unit variance and zero mean. The users maximum power is assumed $P_{maxU} = 20 $dBm and the BS maximum power is $P_{maxBS} = 40 $dBm. We also assume that the bandwidth of each subcarrier is $20$kHz and the noise power spectral density ($N_0$) is $-174 $dBm/Hz. The number of subcarriers is considered as default $N = 8$.  

\begin{figure}[!h]
\centering
\includegraphics[width=5.5in]{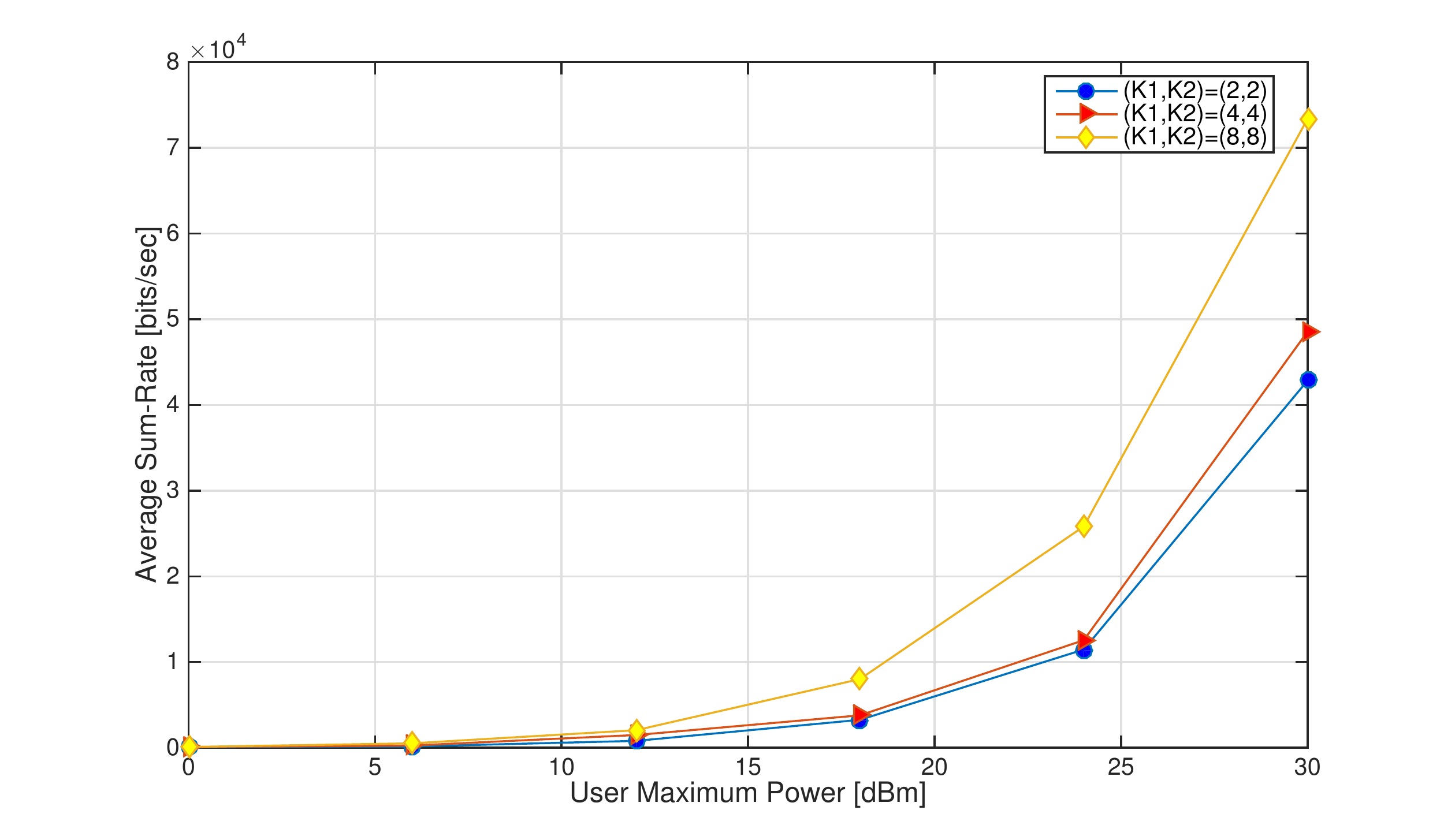}
\caption{Average sum-rate vs. user maximum power with SI effect consideration for the best SINR relay selection scheme.}
\label{f2}
\end{figure}

\begin{figure}[!h]
\centering
\includegraphics[width=5.5in]{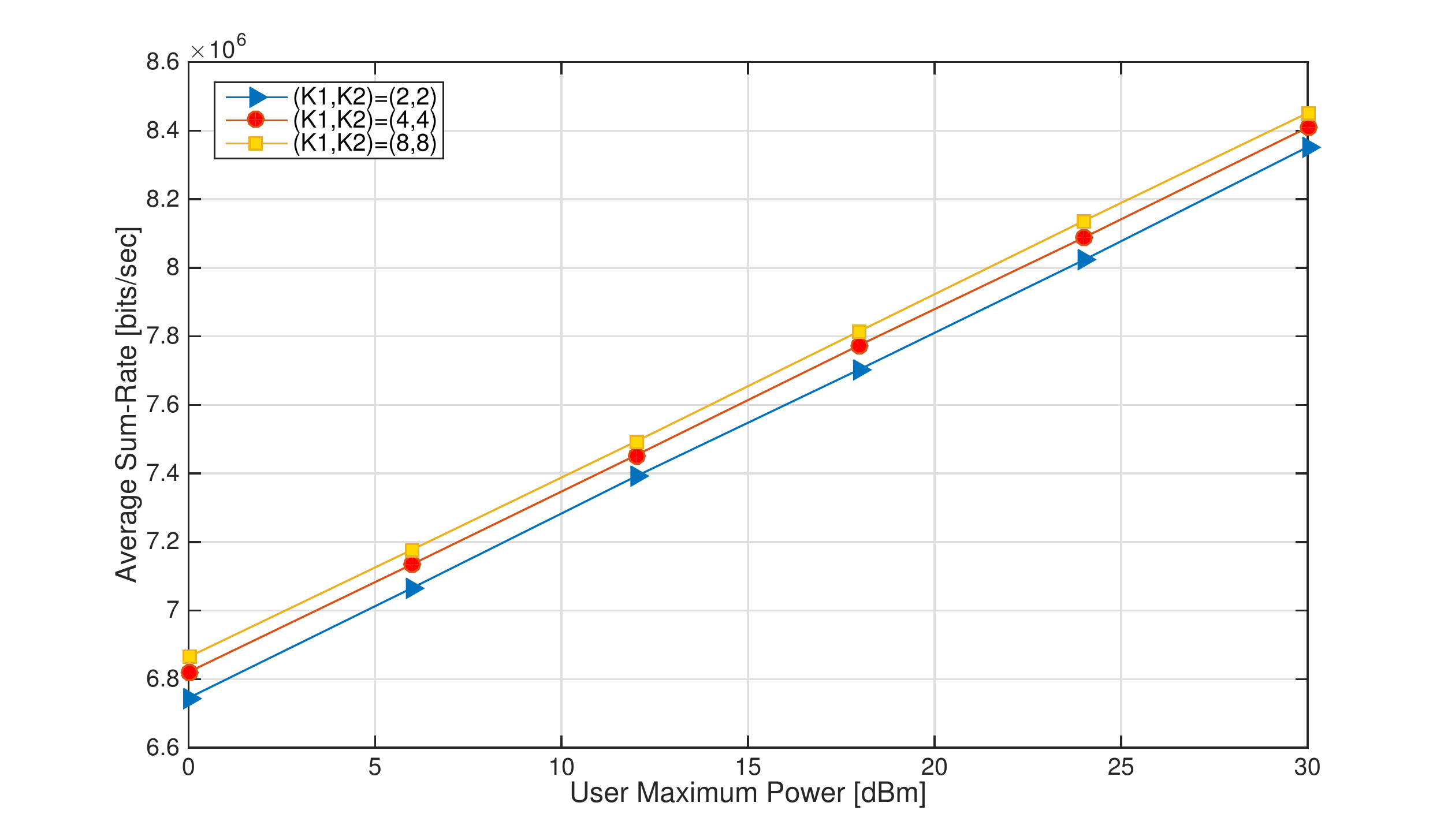}
\caption{Average sum-rate vs. user maximum power without SI effect consideration for the best SINR relay selection scheme.}
\label{f3}
\end{figure}

Fig.$\ref{f2}$ illustrates average sum-rate of users versus different values of user maximum power with consideration of SI effect for different $K1$ and $K2$ values. In this figure, we use the best SINR relay selection scheme with consideration of SI effect and the QoS and power constraints requirements are satisfied. This figure illustrates that when the number of users and relays increases, average sum-rate increases too. Fig.$\ref{f3}$, illustrates the average sum-rate of users versus different values of user maximum power without consideration of SI effect for different $K1$ and $K2$ values. In this figure, also, we use the best SINR relay selection scheme without consideration of SI effect. 

\begin{figure}[!h]
\centering
\includegraphics[width=5.5in]{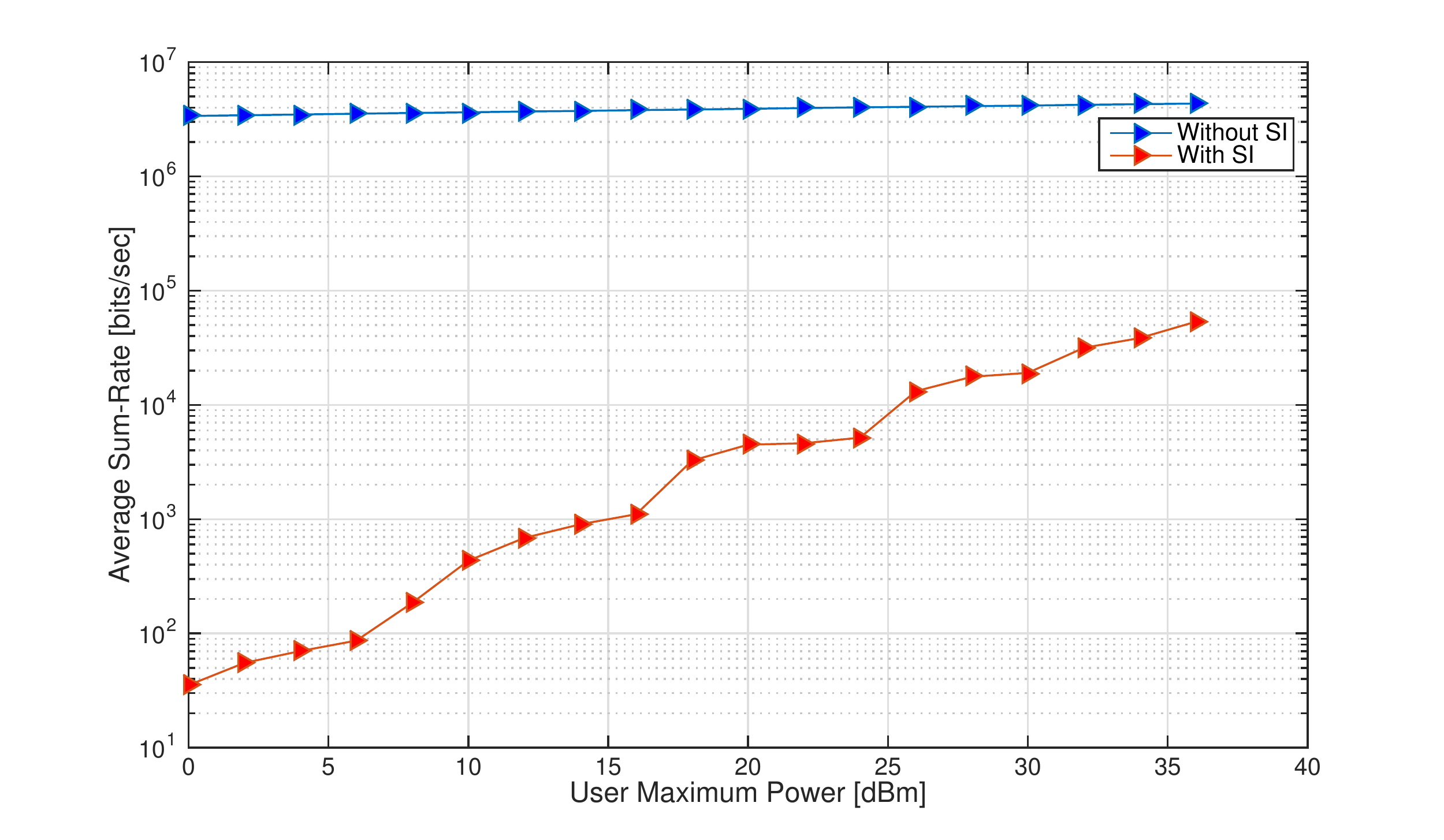}
\caption{Average sum-rate vs. users maximum power for comparison of SI effect for the best SINR relay selection scheme.}
\label{f4}
\end{figure}

In Fig.$\ref{f4}$, average sum-rate versus users maximum power for two cases with best SINR relay selection scheme are presented. In the first case, the BS self-interference effect is considered, while in the second case, it is not considered. It is obvious that when the interference (specifically the BS self-interference) decreases, the sum-rate increases. The self-interference effect is shown clearly in this figure. The number of users and relays is equal to 4.

\begin{figure}[!h]
\centering
\includegraphics[width=5.5in]{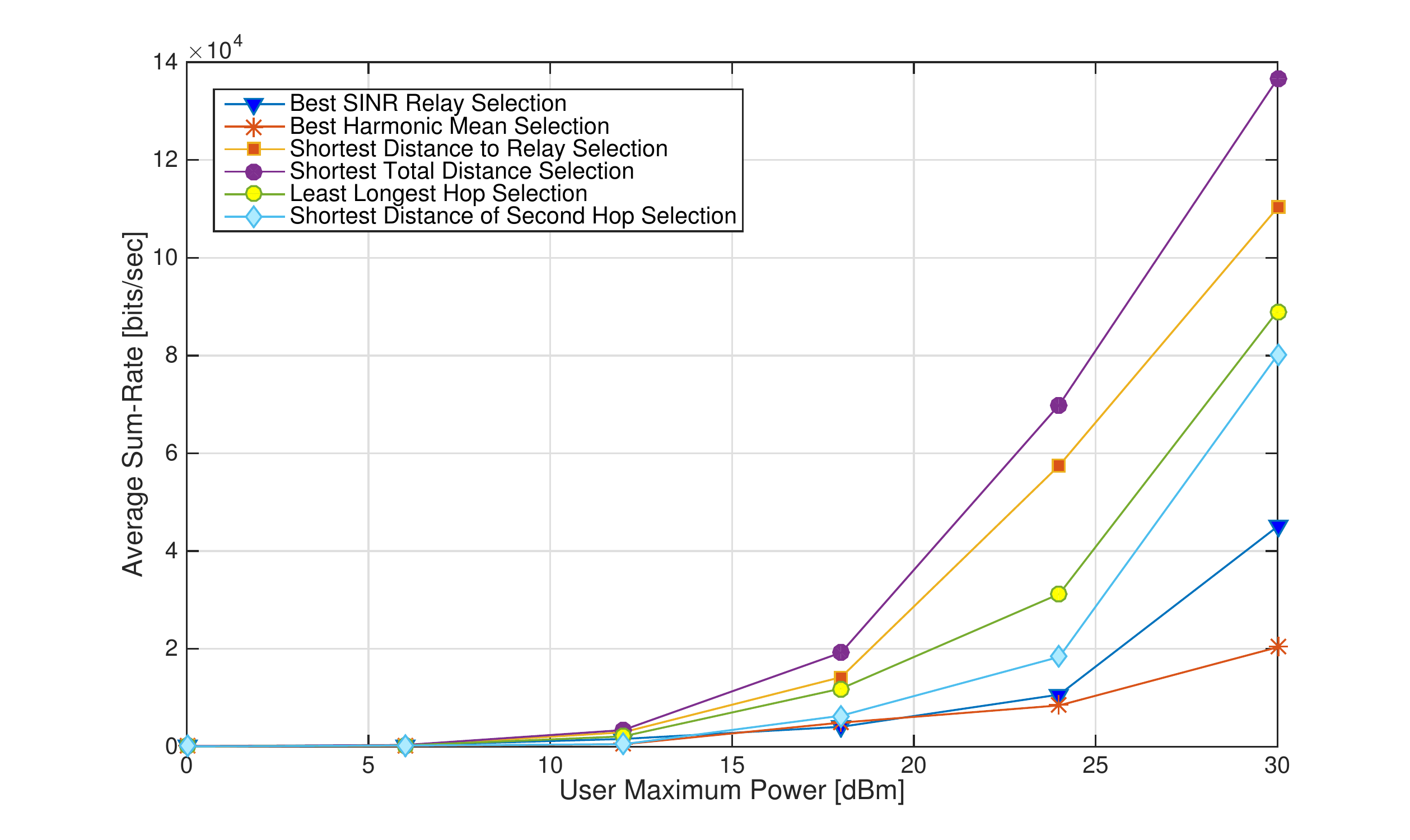}
\caption{Average sum-rate vs. user maximum power with SI effect consideration for the shortest total distance selection scheme.}
\label{f5}
\end{figure}

\begin{figure}[!h]
\centering
\includegraphics[width=5.5in]{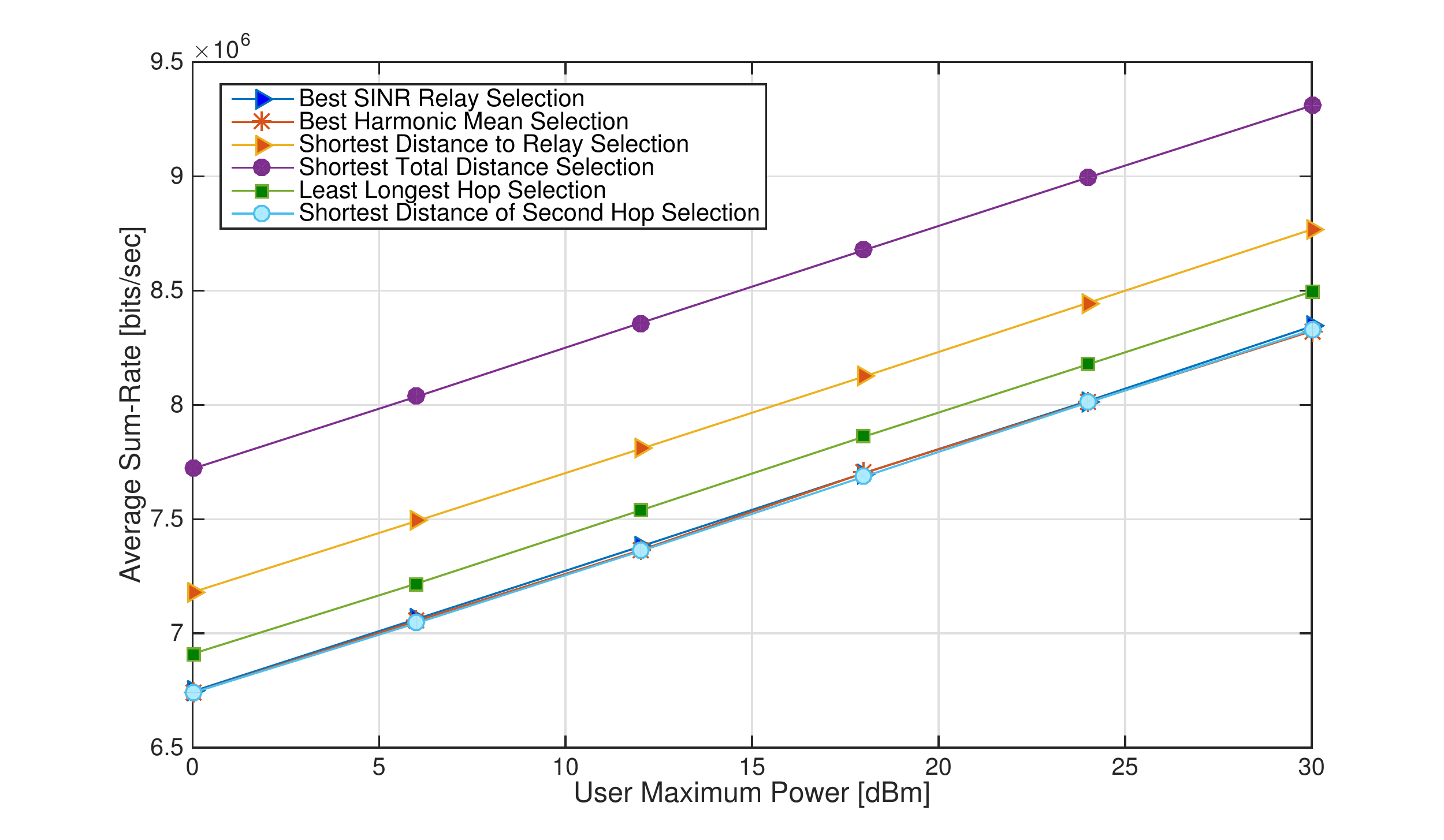}
\caption{Average sum-rate vs. user maximum power without SI effect consideration for the shortest total distance selection scheme.}
\label{f6}
\end{figure}

Fig.$\ref{f5}$ illustrates the average sum-rate of users versus different values of user maximum power with consideration of SI effect to analysis relay selection schemes. The number of users and relays is equal to 4. In this figure, We see that the shortest total distance selection scheme is the best relay selection scheme. It is shown that when the user maximum power is very low, the sum-rate is very low, too and relay selection does not have much effect on the sum-rate. Fig.$\ref{f6}$ is same as Fig.$\ref{f5}$, and illustrates the average sum-rate of users versus different values of user maximum power. This figure shows the effect of relay selection schemes, however, the SI effect is not assumed. The result of this figure is that the shortest total distance selection scheme is the best relay selection method in our scenario. 

\begin{figure}[!h]
\centering
\includegraphics[width=5.5in]{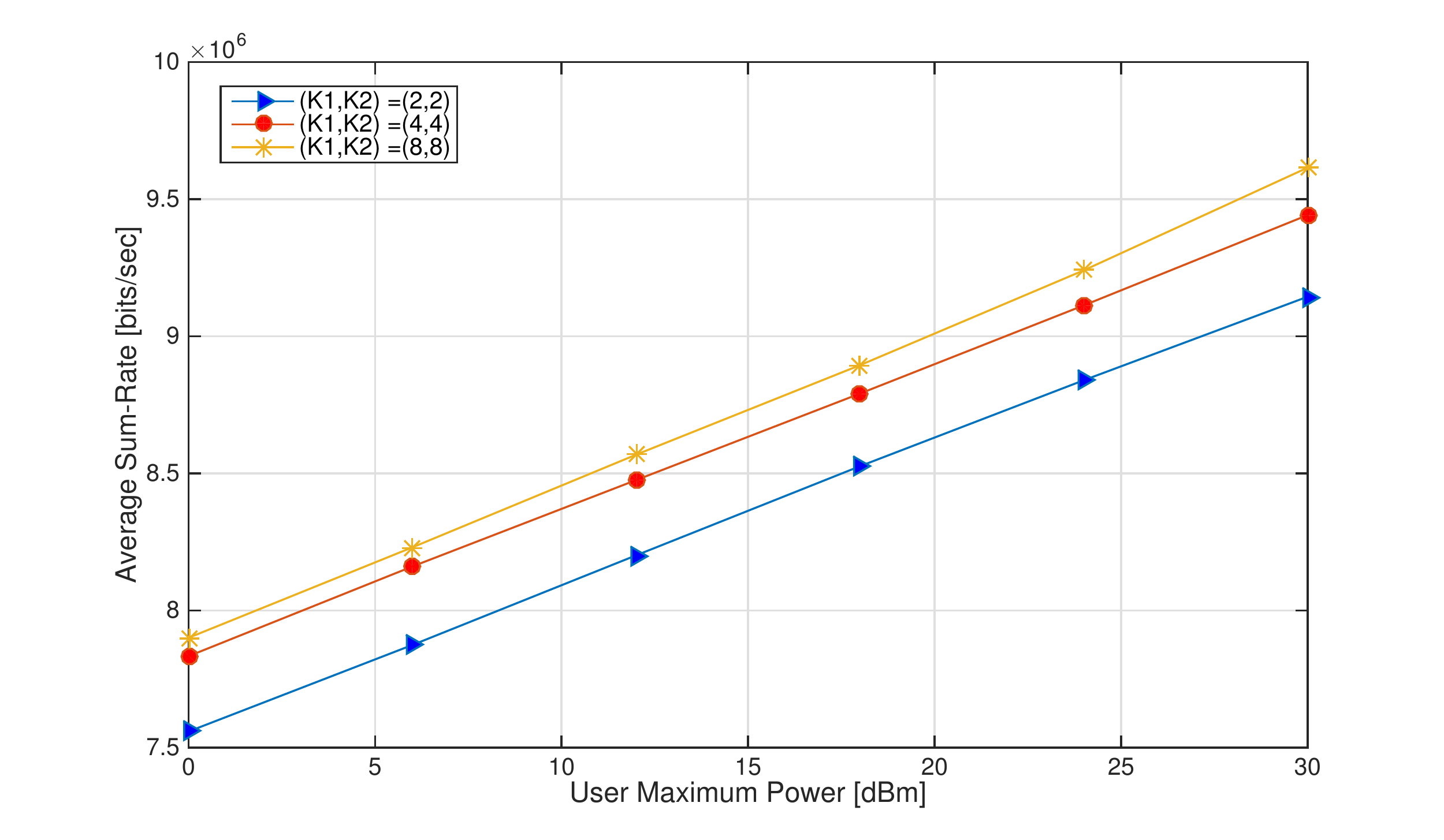}
\caption{Average sum-rate vs. user maximum power without SI effect consideration for the shortest total distance selection scheme.}
\label{f7}
\end{figure}

Fig.$\ref{f7}$ illustrates the average sum-rate versus user maximum power without consideration of SI effect to analyze the effect of different $K1$ and $K2$ values. We applied the shortest total distance selection scheme to the relay selection. This figure illustrates that the number of users and relays have direct effect on the sum-rate. All of our figures satisfies the QoS and power requirements.

\section{Conclusion}\label{S:6}
In this paper, we studied a resource allocation problem and some relay selection techniques in an IBFD cooperative OFDMA system. Our optimization problem is a maximization problem on the network sum-rate with assuming the power and quality of service as constraints. We develop a joint power and subcarrier allocation as well as multiple relay selection schemes to select the best relays to help users with longer distance to the BS. The subcarrier allocation method is a linear assignment algorithm based on Munkres algorithm. In our work, we consider the self-interference in the BS that influences on the data rate. With our simulation results, it is shown that the best relay selection scheme is one that considers the total distance between the user and the BS. Also, simulation results demonstrate that the proposed scheme outperforms the other existing methods.
\section*{References}\label{S:7}

\newif\ifquoteopen
\catcode`\"=\active % lets you define `"` as a macro
\DeclareRobustCommand*{"}{%
   \ifquoteopen
     \quoteopenfalse ''%
   \else
     \quoteopentrue ``%
   \fi
}

\section*{Biography}\label{S:8}
\textbf{Jafar Banar} received his B.Sc. from Amirkabir University of Technology (AUT), Tehran, Iran, in 2013 and his M.Sc. from Iran University of Science and Technology (IUST), Tehran, Iran in 2016, all in Telecommunication engineering. His current research interests include wireless communications and in-band full-duplex. \\

\textbf{S. Mohammad Razavizadeh} is an assistant professor in the school of Electrical Engineering at Iran University of Science and Technology (IUST), Tehran, Iran. His research interests are in the area of signal processing for wireless communications and cellular networks. He is a Senior Member of the IEEE. 

\end{document}